# Spontaneous Ferroelectricity in Strained Low-Temperature Monoclinic $Fe_3O_4$: A First-Principle Study

Xiang Liu, Wenbo Mi[*]

*Tianjin Key Laboratory of Low Dimensional Materials Physics and Processing Technology,*

*School of Science, Tianjin University, Tianjin 300354, China*

---

[*]Author to whom all correspondence should be addressed

E–mail: miwenbo@tju.edu.cn

# ABSTRACT

As a single phase multiferroic material, $Fe_3O_4$ exhibits spontaneous ferroelectric polarization below 38 K. However, the nature of ferroelectricity in $Fe_3O_4$ and external disturbance like strain on it remain ambiguous. Here, the spontaneous ferroelectric polarization of low-temperature monoclinic $Fe_3O_4$ was investigated by first-principles calculations. The pseudo-centrosymmetric Fe$B$42–Fe$B$43 pair has a different valence state. The noncentrosymmetric charge distribution results in the ferroelectric polarization. The initial ferroelectric polarization direction is in the minus direction of $x$ and $z$ axes. The ferroelectricity along $y$ axis is limited due to the symmetry of $Cc$ space group. Both the ionic displacement and charge separation at Fe$B$42–Fe$B$43 pair are affected by strain, which further influences the spontaneous ferroelectric polarization of monoclinic $Fe_3O_4$. The ferroelectric polarization along $z$ axis shows an increase of 45.3% as the strain changes from 6% to -6%.

## I. INTRODUCTION

Recently, the multiferroic materials which combine ferroelectricity, ferromagnetism and ferroelastic characteristics provide a feasibility to realize complex functional devices, so it has attracted much interest because of the ability of changing their magnetic states through electric field and vice versa.[1-3] The ferroelectric polarization switch in multiferroic tunnel junction can realize the multiple-resistance states, which can enhance the storage density.[4-8] Low energy consumption and high response in multiferroic heterostructures provide the potential applications in the next-generation sensors and spintronics devices.[9-12] As a natural magnet, $Fe_3O_4$ has been used for thousands of years. $Fe_3O_4$ is a candidate for spintronics devices owing to its high Curie temperature of 858 K,[13] theoretical spin polarization of -100%[14] and large net magnetic moment of ~4 $\mu_B$/f.u.[15]. A first-order phase transition which is known as Verwey transition has been observed in the temperature-dependent conductivity of $Fe_3O_4$ around 120 K ($T_V$),[16] where the conductivity changes by about two orders of magnitude across $T_V$. Below $T_V$, the unit cell distorts to a $\sqrt{2}a \times \sqrt{2}a \times 2a$ monoclinic superstructure with $Cc$ space group symmetry, where $a$ is the lattice constant of cubic $Fe_3O_4$. Below 38 K, $Fe_3O_4$ shows ferroelectricity[17] (FE) and magnetoelectric coupling at 4.2 K[18]. Yamauchi et al.[19] have studied the spontaneous ferroelectric polarization in $Fe_3O_4$ by first-principles calculations. The ferroelectric polarization along $x$, $y$ and $z$ axes are about -4.41, 0, and 4.12 $\mu C/cm^2$, respectively.[19] However, several experimental results show a significant difference. At 4.2 K, Kato et al.[20] reported a ferroelectric polarization of 1.5 $\mu C/cm^2$ along $[001]_m$, while a large ferroelectric polarization of 52 $\mu C/cm^2$ along $[100]_m$ was reported by Miyamoto et al.[21] Recently, Alexe et al.[17]

investigated the ferroelectricity of $Fe_3O_4$ in metal/$Fe_3O_4$/Nb-doped $SrTiO_3$ heterostructures, which were prepared by pulsed laser deposition (PLD) and magnetron sputtering (MS). Around 10 K, the switchable ferroelectric polarization (2**P**) projected on *x–y* plane is about 11 (PLD) and 5 μC/cm$^2$ (MS). There are two possible explanations for the different experimental results. Firstly, the leakage, antiphase boundaries or residual strain may hamper the ferroelectric polarization measurements.[20] Secondly, the discrepancy can be illustrated by the modern theory of polarization, which recognizes the ferroelectric polarization as a superposition of lattice rather than a vector.[22-25] In this letter, a serious of investigations on the origin and direction of spontaneous ferroelectric polarization and strain effects on it of low-temperature monoclinic $Fe_3O_4$ are performed. Remarkably, the ferroelectric polarization shows an increase of 45.3% as the strain changes from 6% to -6%, which has the potential applications in multiferroic devices.

## II. CALCULATION DETAILS AND MODELS

Electronic structure calculations are performed based on the density functional theory and generalized gradient approximation using Vienna Ab-initio Simulation Package (VASP).[26,27] Projector-augmented-waves method is used to describe the core electrons. The exchange-correlation functional parameterization of Perdew-Burke-Erzenhof is used. The energy cutoff and the Monkhorst-Pack grid of *k*–points are set as 400 eV and 3×3×2, respectively. The on-site Column interaction parameter *U*=4.5 eV and on-site exchange interaction parameter *J*=0.89 eV for Fe ions are used.[28-31] The lattice constants, atomic labels and positions of low-temperature monoclinic Fe$_3$O$_4$ are selected from Ref. 32. The structure optimizations stop until the total energy change is less than 10$^{-5}$ eV and the Hellman-Feynman force change of optimized structure falls below 10$^{-2}$ eV/Å.

Lattice strain ranging from 6% to -6% is defined as $S=(b-a)/a\times100\%$, where $S$, $a$, and $b$ are the strain, lattice constants without and with strain, respectively. Both tensile and compressive strains are imposed along $x$ and $y$ axes. Berry phase is calculated to determine the contribution of ferroelectric polarization from both ions and electrons. In order to distinguish the spontaneous ferroelectric polarization, a paraelectric (PE) structure must be considered because the value of polarization for a PE structure isn't always zero.[25] The PE structure with $C2/c$ symmetry, which is a subgroup of $Cc$ symmetry, is obtained by an online server.[33-35] Several intermediate structures are constructed between FE and PE structure. Although the exact displacements through a ferroelectric polarization switching are more complicated, all of the ions are assumed to move through a linear path. The displacement of ions is scaled by $\lambda$ where $\lambda=100\%$ and $\lambda=0\%$ represent $Cc$ (FE) and $C2/c$ (PE) structures. Negative $\lambda$, i.e., the enantiomorphic counterpart with -($Cc$) structure is also considered. The ferroelectric polarization is calculated with Berry phase, point charge model (PCM) and polarization quantum. The calculations based on PCM are performed by our homemade codes. The calculations of PCM are expressed as $\mathbf{P}=\sum_i q_i \mathbf{r}_i /\Omega$, where $q_i\mathbf{r}_i$ represents the electric moment of the $i^{th}$ ion with $q_i$ charges at $\mathbf{r}_i$, $\Omega$ is the volume of the unit cell. $\mathbf{r}_i$ is treated as a vector from the body center to the $i^{th}$ ion within our codes. Ions at vertex, face center and edge are placed at all of the equivalent sites to eliminate the extra ferroelectric polarization throughout the PCM calculations.

## III. RESULTS AND DISCUSSION

Figure 1(a) shows the lattice structure of low-temperature monoclinic $Fe_3O_4$. It's found that $Fe_3O_4$ with $Cc$ space group has a pseudo-centrosymmetric structure. Thus, the artificial centrosymmetric $Fe_3O_4$ with $C2/c$ space group has a similar structure (not shown here). Firstly, the electronic structure in both FE and PE phase is compared. All of the pseudo-centrosymmetric ion pairs in FE phase, except Fe$B$42–Fe$B$43 pair [in Fig. 1(a)], possess the similar density of states (DOS), which suggests a similar valence state for the counterpart. Figures 1(b)-(d) show the partial DOS projected on $3d$ orbits of Fe$B$42 and Fe$B$43 at $S$=0% and $\lambda$=100%. It is notd that, the ionic sites at Fe$B$42 and Fe$B$43 are equivalent within $C2/c$ symmetry. In Fig. 1(b), DOS at Fe$B$42 (Fe$B$43) does not show a band gap, but the occupation state at Fermi level is zero. The valence of Fe$B$42 (Fe$B$43) with $C2/c$ symmetry is more like +2.5. In FE phase, the $t_{2g}$ orbital with minority-spin state is occupied at Fe$B$42 [Fig. 1(c)] and unoccupied at Fe$B$43 [Fig. 1(d)] at $\lambda$>0%, where the valence of Fe$B$42 and Fe$B$43 is divalent-like and trivalent-like, respectively. At $\lambda$<0%, the occupation states are inverted, which manifests an inversed valence states. Thus, we suppose that the ferroelectric polarization comes from the non-centrosymmetric charge distribution at Fe$B$42–Fe$B$43 pair.

It should be pointed out that the Berry phase results can't confirm the direction of ferroelectric polarization and distinguish the polarization from single ion. In this point of view, PCM is used to calculate the ferroelectric polarization. Here, the results based on PCM are only a rough estimate. Figure 2 shows the dependence of ferroelectric polarization on ionic displacement and charge separation at $S$=0%. The charge separation is defined by the difference of valence between trivalent-like and divalent-like ions. The ferroelectric polarization along $y$ axis is forbidden owing to the symmetry limitation. The ferroelectric polarization increases as charge separation is enhanced, but decreases with the increased ionic displacement. One can see that the low-temperature monoclinic $Fe_3O_4$ has the ferroelectric polarization when the lattice structure loses centrosymmetry. The shining balls in each color map represent the ferroelectric polarization from the Fe*B*42–Fe*B*43 pair. It is hard to distinguish the color of shining balls and the background which demonstrates that the ferroelectric polarization in monoclinic $Fe_3O_4$ almost comes from the Fe*B*42–Fe*B*43 pair. Similar results are also found with other strains (not shown here). We can conclude that once $Fe_3O_4$ loses its centrosymmetry, the equivalent Fe*B*42$^{2.5+}$–Fe*B*43$^{2.5+}$ pair will divide into the divalent-like and trivalent-like ions. The noncentrosymmetric charge distribution of Fe*B*42–Fe*B*43 pair results in the ferroelectric polarization in monoclinic $Fe_3O_4$. The absence of ferroelectric polarization at temperature just below $T_V$ can be ascribed to the low resistance of $Fe_3O_4$, which results in a large current leakage.[17] By PCM calculations, the direction of ferroelectric polarization can be determined. The ferroelectric polarization from other ions can be treated as zero. The ferroelectric polarization direction of Fe*B*42–Fe*B*43 pair is the direction of ferroelectric polarization for the monoclinic $Fe_3O_4$. A vector **R** from body center to the center of Fe*B*42–Fe*B*43 pair is defined as **r**$_{FeB42}$+**r**$_{FeB43}$ and calculated. At $S$=0% and $\lambda$=100%, **R**=(0.128, 0, 0.116) Å, which manifests that the center of Fe*B*42–Fe*B*43 pair is away from the body center along +$x$ and +$z$ direction. However, the ferroelectric polarization is along -$x$ and -$z$ direction due to a higher valence state of Fe*B*43. The initial ferroelectric polarization direction is different from the -$x$ and +$z$ direction, as reported by

Yamauchi *et al.*[17]

Figure 3 shows the dependence of ferroelectric polarization by Berry phase calculations and total energy on ionic displacement at $S=0\%$, $\pm2\%$, $\pm4\%$ and $\pm6\%$. In Fig. 3(a), the ferroelectric polarization of enantiomorphic counterpart structures exhibits the similar values but an inverse direction. The positive (negative) value represents that the direction of ferroelectric polarization is in the plus (minus) direction of coordinates. The opposite value of ferroelectric polarization in enantiomorphic structures comes from the inverse valence state of Fe$B$42–Fe$B$43 in the relevant structures. In Fig. 3(b), Fe$_3$O$_4$ with enantiomorphic structures possesses a similar total energy, which stabilizes at $\lambda=\pm100\%$ except for $S=2\%$. The artificial $C2/c$ structure is unstable based on the calculations. Unlike the results of ferroelectricity in BiFeO$_3$[25], the ferroelectric polarization of Fe$_3$O$_4$ doesn't show a linear-like dependence on displacement. As we know, the polarization in a limited three-dimension system is defined as $\mathbf{P} = \sum_i q_i \mathbf{r}_i / \Omega$, where the ferroelectric polarization shows a linear correlation with the ionic distribution. However, the calculated results seem unreasonable at the first sight. In common perovskite ($AB$O$_3$) ferroelectrics, the ferroelectric polarization comes from the $B$-ion displacement.[25,36,37] Whereas in Fe$_3$O$_4$, the ferroelectric polarization is dominated by charge separation at Fe$B$42–Fe$B$43 pair, namely, the nature of charge ordering in Fe$_3$O$_4$ below $T_V$ rather than ionic displacement.

In order to investigate the strain effects on charge separation at Fe$B$42–Fe$B$43 pair, the bond-valence sum[38-40] (BVS) is calculated to estimate the ionic valence. The $\mathrm{BVS} = \sum_i^n \exp\left[(R_0 - R_i)/b\right]$ of ion is decided by the surrounding bond lengths, where $R_0$ and $R_i$ represent the bond-valence parameter and the $i^{\mathrm{th}}$ bond length, $b$ is a constant of 0.37 Å.[38,40] For $Fe^{2+}$–$O^{2-}$ and $Fe^{3+}$–$O^{2-}$, $R_0$ are 1.734 and 1.759 Å, respectively.[40] Unfortunately, the BVS method over(under)-estimates ionic valence as the compress (tensile) strain is applied. However, the relative valence can still distinguish the $Fe^{2+}$ and $Fe^{3+}$ ions. Figure 4(a) shows the dependence of charge separation on lattice strain at $\lambda$=100%. The charge separation increases when the strain changes from -6% to 4%, but decreases at $S$>4%. The change of charge separation at $S$>4% is consistent with our previous results,[41] which can be attributed to the fluctuation of Fe$B$42–O bonds. The dependence of $x(z)$-component of **R** on strain is also investigated. In Fig. 4(a), the increase of $x(z)$-component with strain manifests the increase and decrease of electric moment from Fe$B$42 and Fe$B$43, respectively. Hence, in Fig. 4(b), the resultant ferroelectric polarization reduces when the strain changes from -6% to 6%. Although the $z$-component of **R** shows a decrease at $S$=6%, the reduced charge separation restrains the increase of ferroelectric polarization.

Based on the modern theory of polarization, the ferroelectric polarization calculated through Berry phase for solid is a superposition of lattice.[42] Along each crystal axis, there is a polarization quantum PQ=$eR/\Omega$, where $e$, $R$ and $\Omega$ are the elementary charge, lattice constant along a specific crystal axis and the volume of the unit cell. The output results from VASP should be divided by the polarization quantum, where the remainder is the original state of ferroelectrics. However, the origin state can't be detected by measurements. In experiments, the spontaneous ferroelectric polarization is usually defined by the half value of ferroelectric polarization difference between the positive and negative bias. Actually, the difference of ferroelectric polarization is basically consistent with the polarization quantum. In this point of view, the spontaneous ferroelectric polarization is calculated based on polarization quantum. Figure 4(b) shows the strain-dependent ferroelectric polarization calculated with Berry phase, PCM and half polarization quantum. The polarization quantum expression shows that the spontaneous ferroelectric polarization is directly correlated with the size of the unit cell. The polarization quantum along $x$ and $z$ axes shows a linear-like correlation with strain. The spontaneous ferroelectric polarization estimated by PCM shows a correspondent trend. However, the Berry phase results show a difference from polarization quantum and PCM results. In facts, the electrons around ions show an off-center distribution because of the Fe–O bonding, which can't be simply evaluated by PCM. Similarly, the polarization quantum doesn't take Fe–O bonding into consideration. In Berry phase calculations, the origin of ferroelectric polarization is divided into ionic and electronic ones by using Wannier functions which are localized in space and have been used to visualize the chemical bonding and calculate the electronic structure.[42] The off-center electronic distribution results in an extra ferroelectric polarization. The discrepancy between Berry phase and other results manifests that the electronic distribution in monoclinic $Fe_3O_4$ has a large influence on ferroelectric polarization. Remarkably, as the strain changes from 6% to -6%, the ferroelectric polarization along $z$ axis in low-temperature monoclinic $Fe_3O_4$ can realize an increase of 45.3%.

## IV. CONCLUSION

In summary, the spontaneous ferroelectric polarization and strain effects on ferroelectricity of the low-temperature monoclinic $Fe_3O_4$ are investigated. It is found that the ferroelectric polarization in $Fe_3O_4$ comes from the non-centrosymmetric charge distribution at Fe$B$42–Fe$B$43 pair. The initial direction of ferroelectric polarization which is decided by PCM calculations is in the minus direction of $x$ and $z$ axes. By importing the biaxial strain along $x$ and $y$ axes, the ferroelectric polarization in both $x$ and $z$ direction are tailored. It is found that both the ionic displacement and charge separation at Fe$B$42–Fe$B$43 pair which further influences the ferroelectric polarization in low-temperature monoclinic Fe$_3$O$_4$ are affected by strain. When the strain changes from 6% to -6%, the ferroelectric polarization along $z$ axis exhibits an increase of 45.3%.

# ACKNOWLEDGMENTS

This work is supported by National Natural Science Foundation of China (51671142 and U1632152), Key Project of Natural Science Foundation of Tianjin City (16JCZDJC37300).

**FIGURE CAPTIONS**

**FIG. 1.** (a) The lattice structure of low-temperature monoclinic $Fe_3O_4$ with $Cc$ space group. Fe$B$42 and Fe$B$43 are colored with blue and yellow, respectively. (b) DOS of Fe$B$42 (Fe$B$43) with $C2/c$ symmetry. The inset shows the magnification of DOS at $-0.2 \leq E \leq 0.2$ eV. (c)-(d) DOS of Fe$B$42 and Fe$B$43 within $Cc$ symmetry.

**FIG. 2.** Ferroelectric polarization along (a) $x$ and (b) $z$ axes calculated with PCM at $S=0\%$. The color scale is shown in the right column. Grey lines indicate the isoline of the ferroelectric polarization.

**FIG. 3.** Dependence of (a) ferroelectric polarization along $x$ and $z$ axes and (b) total energy on ionic displacement at $-6\% \leq S \leq 6\%$.

**FIG. 4.** (a) Dependence of charge separation and $x(z)$-component of **R** of Fe$B$42–Fe$B$43 pair on strain. (b) Spontaneous ferroelectric polarization along $x$ and $z$ axes calculated with Berry phase, PCM and polarization quantum at $-6\% \leq S \leq 6\%$. $\lambda=100\%$ are used for both (a) and (b).

Figure 1

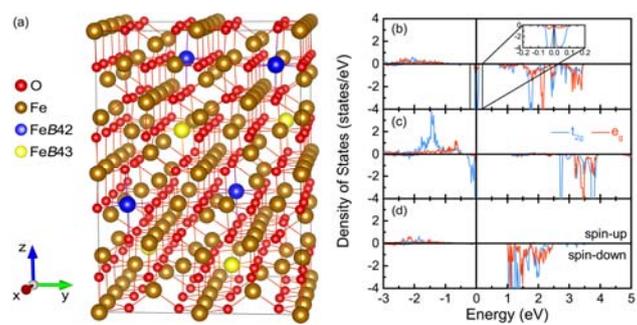

Figure 2

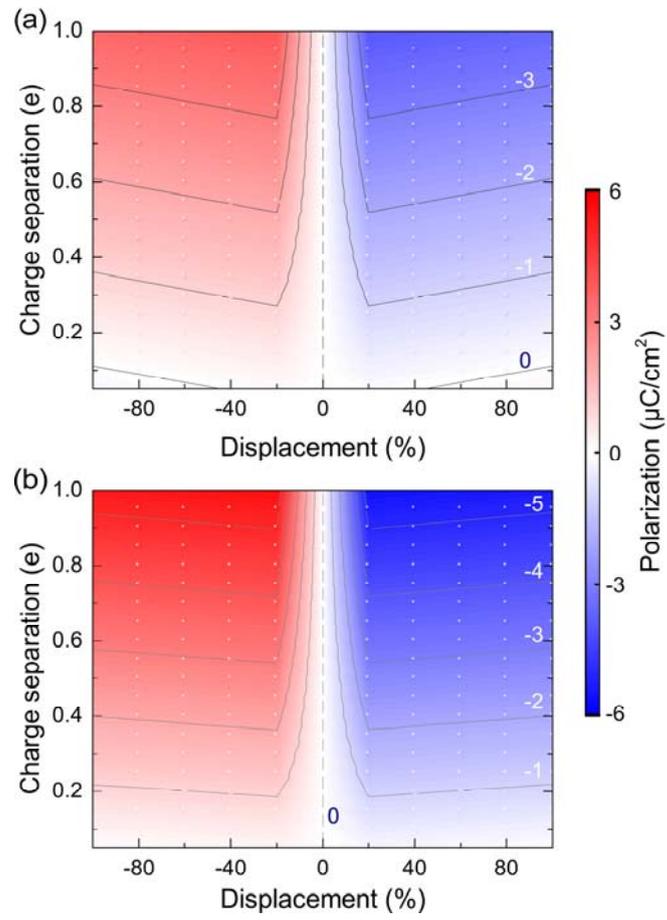

Figure 3

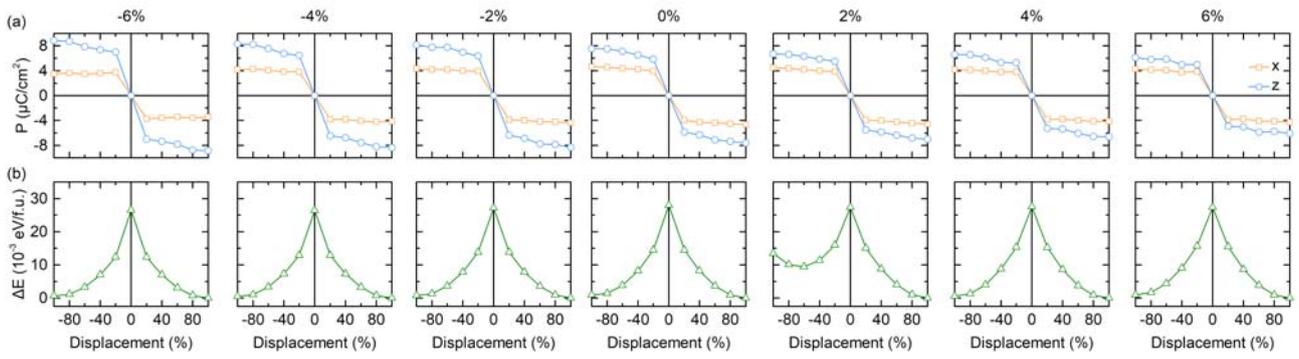

Figure 4

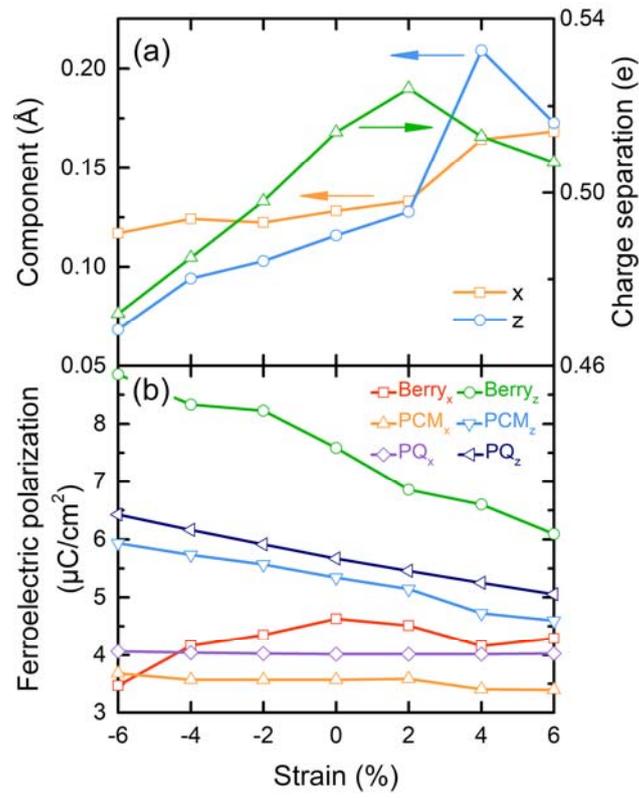